# Multiplex ultrasound imaging of perfluorocarbon nanodroplets enabled by decomposition of post-vaporization dynamics

*Austin Van Namen[1], Sidhartha Jandhyala[1], Catalina-Paula Spatarelu[1], Kenneth M. Tichauer[2], Kimberley S. Samkoe[1,3], and Geoffrey P. Luke[1,3,]\**

[1] Thayer School of Engineering, Dartmouth College, NH, USA

[2] Biomedical Engineering, Illinois Institute of Technology, Chicago, IL, USA

[3] Translational Engineering in Cancer Research Program, Dartmouth Cancer Center, Lebanon, NH, USA

ABSTRACT:

Among the various molecular imaging modalities, ultrasound imaging benefits from its real-time, nonionizing, and cost-effective nature. Despite its benefits, there is a dearth of methods to visualize two or more populations of contrast agents simultaneously – a technique known as multiplex imaging. In this paper, we present a new approach to multiplex ultrasound imaging using perfluorocarbon (PFC) nanodroplets. The nanodroplets, which undergo a liquid-to-gas phase transition in response to an acoustic trigger, act as activatable contrast agents. By using two populations of PFC nanodroplets, each with a different core boiling point, their unique temporal responses to an acoustic trigger were leveraged to differentiate their unique contributions to the overall ultrasound signal. This work characterized the dynamic responses of two PFC nanodroplets with boiling points of 28 and 56 °C. These characteristic responses were then used to demonstrate that the relative concentrations of the two populations of PFC nanodroplets could be accurately measured in the same imaging volume within an average error of 1.1%. Overall, the findings indicate the potential of this approach for multiplex ultrasound imaging, allowing for the visualization of multiple molecular targets simultaneously.

Molecular imaging harnesses targeted contrast agents to provide cell- or molecule-specific contrast.[1] The ability to noninvasively image molecular information has the promise to enable the precise application of highly specific therapeutics. Several molecular imaging techniques have been developed and applied in preclinical and clinical applications, including positron emission tomography,[2] fluorescence imaging,[3] magnetic resonance imaging,[4] photoacoustic imaging,[5] and ultrasound (US) imaging.[6]

Ultrasound imaging has the benefit of being a real-time, nonionizing, and inexpensive imaging modality. Gaseous microbubbles are commonly used as contrast agents for US imaging.[6] The microbubbles typically contain a gaseous perfluorocarbon or oxygen core and a protein or lipid stabilizing shell. Importantly, the microbubbles generate excellent ultrasound contrast; single-microbubble sensitivity can be achieved. This has led to the development of novel imaging techniques, such as super-localization US imaging.[7] Targeting molecules, such as antibodies, can be attached to the surface to confer molecular specificity. Various biological targets, including the vascular endothelial growth factor receptor and the $\alpha_v\beta_3$ integrin have been successfully imaged with US.[8] In spite of the strengths of microbubbles, their large size (typically 1 to 5 μm in diameter) restricts their applicability to intravascular targets. In addition, the relatively low stability of microbubbles limits their lifetime in circulation to a few minutes.[9]

Perfluorocarbon (PFC) nanodroplets have emerged as a promising alternative to microbubbles.[10-11] They contain a liquid (rather than gaseous) PFC core. Because liquid PFCs have similar mechanical properties to tissue, they provide negligible inherent contrast. They can, however, be activated by a burst of acoustic or optical energy to undergo a liquid-to-gas phase transition, a process known as acoustic droplet vaporization (ADV)[12] or optical droplet vaporization (ODV),[13] respectively. The resulting gaseous bubbles can be detected with US imaging with single bubble sensitivity. Depending on the ambient temperature and the boiling point of the PFC, the bubbles will either persist (low boiling point) or recondense back into their nanodroplet form (high boiling point), primed to undergo another vaporization event.[14-16]

Nanodroplets have many strengths over their microbubble counterparts. First, their small size and liquid core make them more stable in biological conditions and more likely to reach extravascular targets.[17] Second, they can be activated on-demand with externally applied energy.[18] This gives an extra level of control that can be leveraged to boost US contrast.[14] Third, they can be loaded with optical dyes, nanoparticles, or therapeutic molecules for multimodal imaging or image-guided drug delivery.[13, 19-21]

Multiplex imaging (the ability to detect and distinguish between two different formulations of contrast agents in the same imaging volume) expands the utility of molecular imaging to simultaneously visualize two markers or to control for nonspecific accumulation and binding.[22-23] Multiplex US imaging has proven elusive thus far. In theory, the resonant frequency of microbubble could be tuned by varying their size and shell composition.[24] In practice, the polydispersity in their size makes it difficult to achieve two distinct populations of microbubbles. Alternatively, optical absorbers with distinct absorption spectra have been loaded in PFC nanodroplets.[25] Then, a tunable laser can be used to activate only a single population of nanodroplets at a time. The high attenuation of light in tissue, however, limits the imaging depth that can be achieved.

In this paper, we describe a new approach to multiplex US imaging. We synthesized two formulations of PFC nanodroplets: one with a perfluoropentane core (bulk boiling point = 28 °C) and the other with a perfluorohexane core (boiling point = 56 °C). Both populations of nanodroplets can be vaporized with the same pulse of focused ultrasound (FUS) energy. The perfluoropentane nanodroplets ($ND_{28}$) undergo a single vaporization event, whereas the perfluorohexane nanodroplets ($ND_{56}$) recondense after forming a transient bubble.[16] This distinct behavior was harnessed to develop an imaging strategy that can effectively differentiate between the two populations of nanodroplets. This work demonstrates that the relative ratio of the concentration of $ND_{28}$ and $ND_{56}$ can be imaged. Overall, this approach could be applied to enable US imaging of multiple biomarkers in the same imaging volume.

**RESULTS AND DISCUSSION**

**Nanodroplet synthesis and characterization**

The perfluorocarbon nanodroplets were synthesized (Fig. 1a) by first forming a lipid cake from a solution containing a mixture of 1,2-dipalmitoyl-*sn*-glycero-3-phosphocholine (DPPC) and 1,2-distearoyl-*sn*-glycero-3-phosphoethanolamine-N-[(polyethylene glycol)-2000] (DSPE-PEG) in chloroform with a 90:10 weight ratio using a rotary evaporator. The lipids were rehydrated with deionized water and dispersed with a water bath sonicator. Then, either perfluoropentane or perfluorohexane was added, and a microtip probe sonicator was applied to produce $ND_{28}$ or $ND_{56}$, respectively. Excess lipids were removed via centrifugation washing.

The nanodroplet size was measured via dynamic light scattering (DLS, Fig. 1b). The $ND_{28}$ had an average peak diameter of $340 \pm 43$ nm; the $ND_{56}$ had an average peak diameter of $350 \pm 47$ nm. The polydispersity index of the nanodroplets was $0.30 \pm 0.11$ and $0.27 \pm 0.07$ for the $ND_{28}$ and $ND_{56}$, respectively. The DLS instrument estimated the concentration of the two samples to be $2.4 \times 10^8$ $ND_{28}$/mL and $2.1 \times 10^8$ $ND_{56}$/mL.

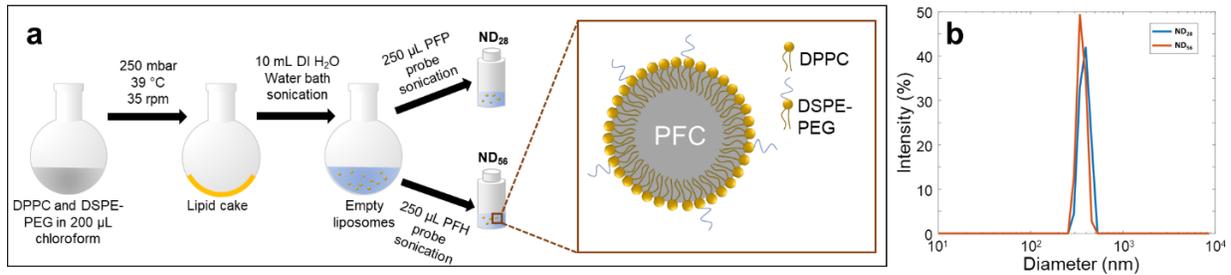

**Figure 1. a)** The synthesis procedure of $ND_{28}$ and $ND_{56}$ relied on sonication to generate a perfluorocarbon-in-water emulsion with phospholipids forming the stabilizing shell. **b**) A histogram of the hydrodynamic diameter showed an average peak of $340 \pm 43$ nm for the $ND_{28}$ and $350 \pm 47$ nm for the $ND_{56}$.

**Nanodroplet dynamics**

    A custom imaging setup was designed that incorporates a 15-MHz linear array ultrasound imaging transducer and a 1.1-MHz single-element focused ultrasound (FUS) transducer (Fig. 2a). The imaging transducer was aligned with the focus of the FUS transducer using a 3-D printed stage. The FUS transducer was coupled to the sample with a custom-molded polyacrylamide coupling cone with ultrasound gel applied to all interfaces. A polyacrylamide gel phantom containing a homogenous mixture of nanodroplets was used as the imaging medium. The FUS and imaging transducers were triggered by a function generator to allow sub-microsecond synchronization between the FUS transmission and US image acquisition.

Three polyacrylamide phantoms containing a homogenous distribution of either $ND_{28}$ or $ND_{26}$ were constructed with a concentration of $1.0 \times 10^6$ nanodroplets/mL. The phantoms were imaged to measure the temporal dynamics of the US signal in response to the FUS activation. An initial increase in ultrasound signal was observed for both $ND_{28}$ and $ND_{56}$ in response to the 1.1-MHz, 10-cycle FUS stimulus (Fig. 2 b-i). In the ensuing US frames, however, the $ND_{28}$ signal gradually increased (Fig. 2 b-e), while the $ND_{56}$ signal decayed back to the baseline (Fig. 2 f-i). The increasing $ND_{28}$ signal is likely attributable to coalescence of the resulting bubbles.[26] The decaying $ND_{56}$ signal is attributable to the recondensation of the nanodroplets after their initial vaporization.[14-16] Our group has previously shown that $ND_{56}$ can be revaporized hundreds to thousands of times with repeated FUS stimuli.[16]

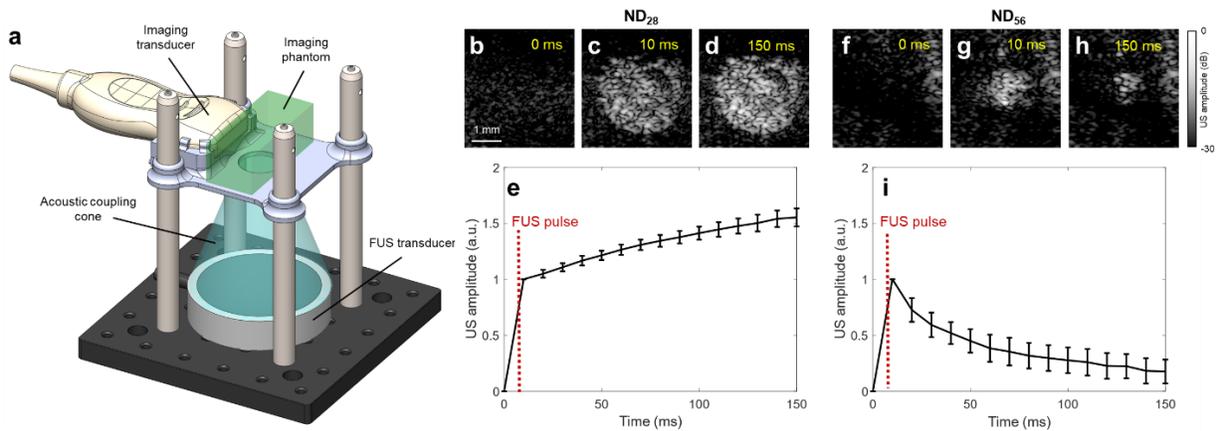

**Figure 2. a)** The experimental setup allowed for colocalized FUS focus and imaging field of view. Ultrasound images of polyacrylamide phantoms containing $ND_{28}$ at **b)** 0 ms, **c)** 10 ms, and **d)** 150 ms show a sustained increase in signal in response to an FUS pulse applied just before 10 ms. **e)** The US signal amplitude shows consistent behavior across n = 8 acquisitions. Ultrasound images of polyacrylamide phantoms containing $ND_{56}$ at **f)** 0 ms, **g)** 10 ms, and **h)** 150 ms show an initial increase in signal in response to the FUS pulse, followed by a gradual decay. **i)** The US signal

amplitude shows consistent behavior across n = 8 acquisitions. Error bars represent the standard deviation.

**Multiplex imaging acquisition optimization**

The US images of $ND_{28}$ and $ND_{56}$ (Fig. 2) demonstrated that the two populations of nanodroplets can exhibit distinct temporal behavior in response to an FUS stimulus. Leveraging these unique responses, an image acquisition strategy was developed to differentiate relative concentrations of mixed $ND_{28}$, $ND_{56}$, and background tissue signals. In order to do so, a matrix with three rows was constructed; each row contained the characteristic dynamic responses of $ND_{28}$, $ND_{56}$, and background tissue to an FUS stimuli (Fig. 3a). The number and spacing of FUS pulses were varied and the product of the singular values of the matrix containing the idealized signals was used as a metric for its invertibility and, thus, the differentiability of the three signals. The product of singular values was maximized for a longer delay between FUS pulses, $\tau_{FUS}$ (Fig. 3b). This allowed for the $ND_{56}$ to decay to near baseline signal before a new vaporization pulse was applied. In addition, increasing the number of FUS pulses beyond 5 yielded diminishing returns (Fig. 3c). Based on the results of this optimization, the image acquisition sequence that used a $\tau_{FUS}$ of 400 ms, and 5 total FUS pulses, was selected.

Next, the number of US images needed was reduced as far as possible while preserving the differentiability of the three signals. This was done using an algorithm adapted from our previous work in wavelength selection for spectroscopic photoacoustic imaging.[27] Briefly, a single column (representing a single US frame) was removed from the matrix containing the three signals. Then, the product of singular values was calculated. This was repeated for each possible frame. The frame which led to the largest product of singular values when removed was deemed to be the least important and discarded. The process was repeated until only 16 US frames remained. This led to the image acquisition strategy shown in Fig. 3d. First, 6 baseline frames were acquired, then 5 FUS pulses were applied, with a pair of US frames acquired 0.5 ms and 400 ms after each FUS pulse. This image acquisition sequence was used for all subsequent experiments.

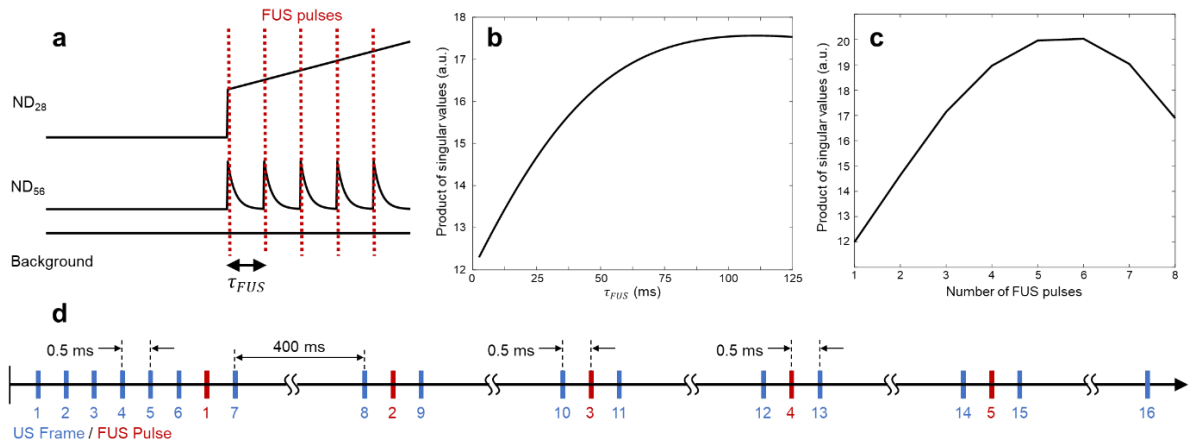

**Figure 3. a)** Simulated time traces for the $ND_{28}$ (top), $ND_{56}$ (middle), and background (bottom). The number and spacing of FUS pulses were tuned to maximize the orthogonality of the three signals. The product of singular values was used as a metric to evaluate the relative differentiability of ND and background signal with respect to **b)** the spacing between FUS pulses $\tau_{FUS}$, and **c)** the number of FUS pulses. **d)** Based on the results from (a-b), an image acquisition strategy consisting of 6 baseline frames and 5 FUS pulses was designed for multiplex imaging.

**Determination of ground-truth signals**

Polyacrylamide phantoms containing a homogenous distribution of either $ND_{28}$ or $ND_{56}$ were imaged using the developed imaging sequence. The concentration of $ND_{28}$ and $ND_{56}$ were matched using DLS measurements. Representative images of the two populations of nanodroplets qualitatively demonstrated that they can be differentiable using the image acquisition strategy (Fig. 4 a-b). The images capture the strong one-time vaporization that $ND_{28}$ exhibited in response to the first FUS pulse, followed by the gradual increase in signal. The $ND_{56}$ exhibited a vaporization signal in response to each of the five FUS pulses, with recondensation occurring before each subsequent FUS pulse was applied. Images from 7 phantoms were acquired to obtain the ground-truth signals of $ND_{28}$ and $ND_{56}$ for the developed imaging sequence (Fig. 4 c-d). These signals were then used for the multiplex imaging studies.

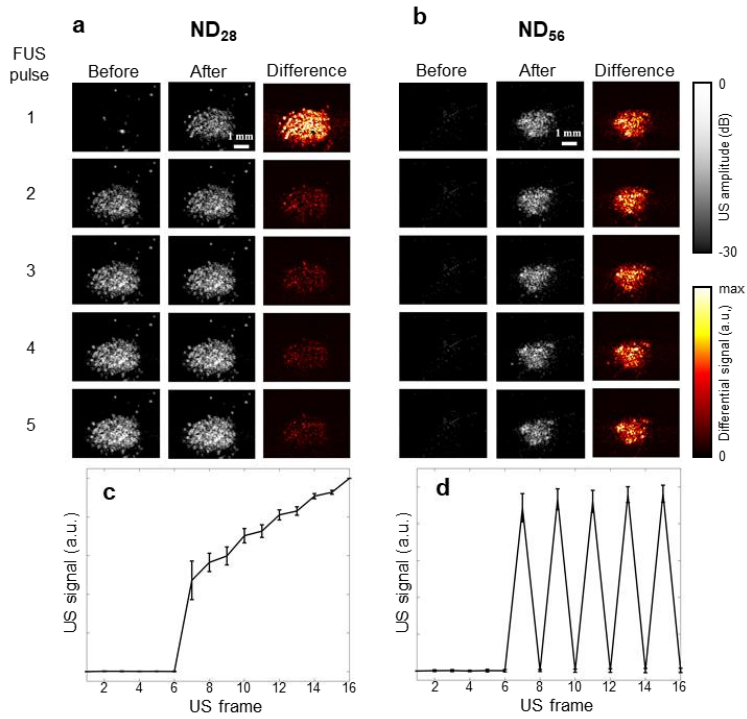

**Figure 4.** Representative US images of phantoms containing **a)** $ND_{28}$ and **b)** $ND_{56}$ before (left column), and 0.5 ms after (middle column) each of the 5 FUS pulses. The difference image (right

column) demonstrates a single large vaporization event for the $ND_{28}$ and multiple vaporization events for the $ND_{56}$. The signals were acquired from 7 phantoms and the average US signal was plotted for **c**) $ND_{28}$ and **d**) $ND_{56}$. The error bars represent one standard deviation.

**Validation of multiplex imaging**

Polyacrylamide phantoms containing homogeneous mixtures of $ND_{28}$ and $ND_{56}$ with a relative percentage of $ND_{56}$ ranging from 0% to 100% in 10% increments were constructed. Five locations in each phantom were imaged using the developed imaging sequence (Fig. 3 d). The signals in a region of interest corresponding to the FUS focal spot were averaged for each US frame. Then, each of these 16-sample ultrasound amplitudes was unmixed using nonnegative least squares to obtain relative contributions of the $ND_{28}$, $ND_{56}$, and background to the US signal. Finally, the relative percentage of $ND_{56}$ was calculated.

The resulting estimate of % of $ND_{56}$ showed a good agreement with the ground truth (Fig. 5 a). The average error in relative $ND_{56}$ concentration was 9.1%, with larger errors occurring at higher $ND_{56}$ concentrations. A linear regression demonstrated high linearity of the estimated relative concentration against the ground truth ($R^2 = 0.996$). It is important to note, however, that the slope of the line was not 0.76, indicating an underestimation of the $ND_{56}$ in these samples. This discrepancy is likely attributable to the fact that the nanodroplet concentrations were measured using DLS, which is not the most accurate method for quantifying nanoparticle concentrations. In future studies, the concentration of the two nanodroplet populations could be matched using their US signal, enabling more accurate unmixing. In addition, imaging noise and artifacts arising from microbubble shadowing could have impacted the results.

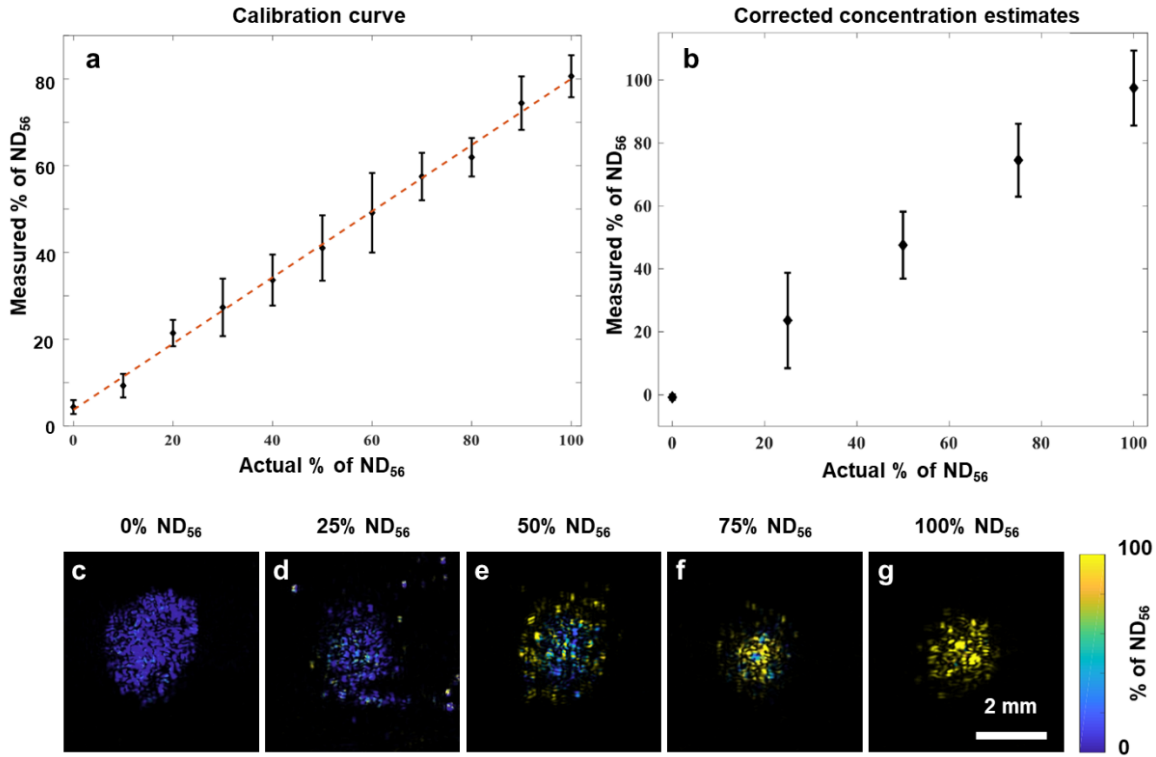

**Figure 5. a**) Average relative concentration of $ND_{56}$ estimated from linear unmixing of the $ND_{28}$, $ND_{56}$, and background signals. Error bars indicate the standard deviation of 5 imaging locations in each of 3 phantoms at each concentration level. The red line indicates a linear fit ($R^2 = 0.996$). **b**) A separate set of phantoms was imaged using the linear fit from (a) as a correction, showing better agreement with the ground truth. Error bars represent the standard deviation of 5 imaging locations for a single phantom at each concentration. Representative images of the corrected unmixed signals show the spatial distribution of the $ND_{28}$ (blue) and $ND_{56}$ (yellow) for phantoms containing a relative concentration of **c**) 0%, **d**) 25%, **e**) 50%, **f**) 75%, and **g**) 100% $ND_{56}$.

In order to correct for the non-ideal concentration estimate, the linear fit in Fig. 5 a was used as a calibration curve for a second set of experiments. Phantoms containing 0%, 25%, 50%, 75%, or 100% ND$_{56}$ were imaged. The estimated concentrations were then scaled by the linear fit from the preceding experiment. The results demonstrated that a much better agreement between the actual and estimated relative ND$_{56}$ concentration with an average error of 1.1%, (Fig. 5 b). In addition, since the processing was carried out in a pixel-wise manner, images of ND$_{28}$ and ND$_{56}$ concentration can be obtained. Representative images of the relative ND$_{56}$ concentration (Fig. 5 c-g) showed that the two populations of nanodroplets can be simultaneously visualized in the same imaging volume. This opens the door to visualizing multiple molecular targets simultaneously in the same imaging volume. It also allows for the possibility of using a non-targeted nanodroplet to help distinguish between contrast agent delivery and specific binding to the molecule of interest. This approach has been successfully applied in fluorescence imaging for highly sensitive detection of a molecule of interest.[28-29]

**CONCLUSIONS**

In this paper, a new multiplex ultrasound imaging strategy was described and initially tested, one that leverages the unique temporal dynamics of two populations of perfluorocarbon nanodroplets in response to the same FUS stimulus. The distinct behavior of the two populations was characterized and used to develop an image acquisition strategy designed to optimize the separation of the ND signals. Finally, the ability to simultaneously visualize the two different nanodroplets in the same imaging volume was demonstrated using tissue-mimicking phantoms. Future work will explore adding molecular targeting to the nanodroplets and applying the technique *in vivo*.

**METHODS**

**Nanodroplet synthesis and characterization**

The overall synthesis procedure is depicted in Fig. 1 a. First, a mixture of 25-mg/mL 1,2-dipalmitoyl-*sn*-glycero-3-phosphocholine (DPPC; Avanti Polar Lipids, Alabaster, AL, USA) in chloroform and 25 mg/mL 1,2-distearoyl-*sn*-glycero-3-phosphoethanolamine-N-[(polyethylene glycol)-2000] (DSPE-PEG; Avanti Polar Lipids) in chloroform was produced with a 90:10 volume ratio of DPPC to DSPE-PEG. Then, 200 µL of the mixture was added to a 50-mL round bottom flask, along with 1 mL of chloroform (Oakwood Chemical, Estill, SC, USA). The chloroform was evaporated using a rotary evaporator (Hei-VAP Precision, Heidolph, Schwabach, Germany) operating at 35 rpm, 250 mbar, and 39 °C. After complete evaporation, the lipid cake was resuspended in 10 mL of deionized water and sonicated for 1 min using a water bath sonicator (Symphony, VWR, Radnor, PA, USA).

The resulting solution was split in half, with 5 mL going into each of two 20-mL glass scintillation vials. One vial was used for the synthesis of perfluoropentane nanodroplets ($ND_{28}$), while the other was used for the synthesis of perfluorohexane nanodroplets ($ND_{56}$). Next, 250 µL of either perfluoropentane or perfluorohexane (Fluoromed, Round Rock, TX, USA) was added to form $ND_{28}$ or $ND_{56}$, respectively. A nanoemulsion was produced in the resulting solutions by sonication with a 3.2-mm tip probe sonicator (Q700, QSonica, Newton, CT, USA). The scintillation vials were placed in an ice bath and sonication was applied at 3 W until 30 J total energy had been applied. The resulting solutions became opaque as the emulsions formed.

The $ND_{28}$ and $ND_{56}$ were isolated using centrifugation (MiniSpin, Eppendorf, Hamburg, Germany). First, the nanodroplets were centrifuged at 43×g RCF for 1 min. The pellet was discarded to remove large particles and lipid aggregates. Then, the samples were washed at 453×g and 4293×g, retaining the pellet in both cases. The final pellet was resuspended in 5 mL of deionized water. This solution served as a stock solution for all experiments.

The nanodroplets were characterized using dynamic light scattering (DLS; Zetasizer Nano ZS, Malvern Instruments, Westborough, MA, USA). The nanodroplets were diluted 67× in deionized water. 1 mL of the diluted sample was transferred into a plastic cuvette with a 1-cm path length. Three measurements with 12 recordings each were performed at 25 °C on each sample. A total of three samples were measured for each nanodroplet. The average peak diameter, polydispersity index, and nanodroplet concentration were determined by the Zetasizer Nano Software.

**Tissue-mimicking phantom preparation**

Polyacrylamide phantoms were used because of their low acoustic impedance, similar mechanical properties to water-based tissue, and room-temperature construction.[30] A 250-mL glass beaker was placed on a stir plate operating at 100 rpm in a fume hood. 66.7 mL of acrylamide/bis-acrylamide 29:1, 30% (Sigma, St. Louis, MO, USA), and 33.3 mL of deionized water were added to the beaker. Then, 100 mg of ammonium persulfate (Sigma, St. Louis, MO, USA) was added to the beaker and allowed to dissolve. The solution was degassed in a water bath sonicator (Symphony, VWR, Radnor, PA, USA) for 30 seconds to eliminate air bubbles, then returned to the stir plate. Then, 500 µL of the stock $ND_{28}$ or $ND_{56}$ solution (or a 500-µL mixture of the two) was added. Finally, 110 µL of N,N,N',N' – tetramethyl-ethylenediamine (TEMED; Sigma, St. Louis, MO, USA) was added to catalyze cross-linking of the hydrogel. The phantom solution was poured into 3-D printed (Mini, Lulzbot, Fargo, ND, USA) rectangular prism molds with dimensions of 7×3×2.5 cm. This allowed for 6-8 distinct measurements to be performed along the length of the phantom during experiments. The solution was allowed to set in a -80 °C freezer for 10 minutes to counteract the effects of the exothermic reaction.

**Ultrasound activation and imaging**

A custom imaging setup was designed to colocalize the FUS focus and the US imaging transducer field of view (Fig. 2 a). A 3-D printed (Mini, Lulzbot, Fargo, ND, USA) stage was designed to hold the imaging transducer rigidly in place. The imaging transducer was a 15-MHz capacitive micromachined linear array (L22-8v, Verasonics, Kirkland, WA, USA). The transducer was powered by a programmable ultrasound imaging system (Vantage 256, Verasonics, Kirkland, WA, USA). Plane wave transmission with angular compounding using 5 angles between -18° and +18° was used for image capture with a 38-μs delay between acquisitions. Image reconstruction was performed using the Verasonics software.

The 1.1-MHz FUS transducer (H-151, Sonic Concepts, Bothell, WA, USA) was powered by a radiofrequency power amplifier (1020L, Electronics & Innovation, Rochester, NY, USA). The FUS waveform was generated by an arbitrary waveform generator (AFG1062, Tektronix, Beaverton, OR, USA), which was triggered by the Verasonics ultrasound system. A 10-cycle sinusoid with peak negative pressure of 11.8 MPa was applied by the FUS transducer to ensure activation of both populations of nanodroplets. These parameters follow our previous characterization of activation thresholds.[16] The FUS transducer was coupled to the tissue-mimicking phantom using a custom-molded polyacrylamide coupling cone. The cone was constructed using the same recipe as the phantoms, but with nanodroplets omitted. Ultrasound coupling gel was applied to all interfaces to facilitate acoustic transmission.

Images of polyacrylamide phantoms containing either $ND_{28}$ or $ND_{56}$ were acquired to characterize the temporal response of each population of nanodroplet. One US image was acquired, followed by the FUS activation pulse, followed by 15 additional US image acquisitions spaced by 10 ms. The FUS focus was identified in the US images by the region of signal increase after the FUS pulse. A 3.9×3.9-mm square region of interest containing the FUS focus was used for all image analysis. The average signal US signal inside this region of interest was used to generate the frame-by-frame plots of US amplitude. The mean and standard deviation of the US amplitude were calculated for n=8 measurements for each population of nanodroplet.

**Imaging sequence simulation and optimization**

In order to optimize the multiplex image acquisition sequence, we sought to maximize the differentiability of the $ND_{28}$, $ND_{56}$, and constant background signals. Based on the results from imaging single populations of nanodroplets, we determined idealized signals for $ND_{28}$ and $ND_{56}$. The $ND_{28}$ were simulated as a step function corresponding to the first FUS pulse, followed by a linear ramp. The $ND_{56}$ were simulated as a sum of decaying exponential functions, with a peak occurring at each FUS pulse. A matrix with three rows was constructed containing idealized amplitude values for $ND_{28}$ (top row), $ND_{56}$ (middle row), and constant background (bottom row). We used the product of the three singular values of this matrix as a metric to determine how differentiable the three different signals would be in practice. Based on this framework, we varied the number of FUS pulses as well as their spacing in time. For each specific FUS characteristics, a new matrix was constructed and the product of singular values was calculated. Parameters that maximized the product of singular values were used for future imaging sequences.

After determining appropriate FUS pulse numbers and spacing, we sought to minimize the number of US images required to unmix the three signals. Again, we used the product of singular values as a metric for signal differentiability. We adapted our previously developed algorithm for wavelength selection in photoacoustic imaging.[31] The method works by iteratively removing a single column from the matrix containing the signals for $ND_{28}$. $ND_{56}$, and background. After removing the column, the product of singular values was calculated on the truncated matrix. The column whose removal resulted in the largest product of singular values was deemed to be the least important and was permanently removed. This process was repeated until only 16 columns (i.e., US images) remained.

Subsequent experiments used the optimized image acquisition sequence (Fig. 3 d). This consisted of first acquiring 6 baseline US images with an inter-frame spacing of 0.5 ms. Then, 5 FUS pulses were applied to the sample. A pair of US images were acquired 0.5 ms and 400 ms after each FUS pulse, resulting in a total of 16 US images.

We constructed phantoms containing only $ND_{28}$ or $ND_{56}$ in order to obtain ground-truth signals for multiplex imaging. The phantoms were imaged with the optimized imaging sequence. The average signal in a 3.9×3.9-mm square region of interest was measured across n=7 phantoms. Finally, a 16×3 matrix was constructed using these normalized signals, with the third row (corresponding to the background) consisting of all ones.

**Multiplex imaging**

Phantoms containing mixtures of $ND_{28}$ and $ND_{56}$ were constructed. The relative percentage of $ND_{56}$ ranged from 0% to 100% in increments of 10%. Three phantoms were constructed for each relative percentage of $ND_{56}$. Five sets of images were acquired in each phantom using the optimized image acquisition sequence. This resulted in 15 measurements for each relative percentage of $ND_{56}$.

The average US signal in the 3.9×3.9-mm region of interest was used for signal unmixing. The relative concentrations of the $ND_{28}$, $ND_{56}$, and background were calculated using nonnegative least squares optimization (Mathworks Matlab) with the measured US signal and the previously constructed ground-truth matrix. The relative concentration percentage of $ND_{56}$ was calculated by:

$$\% \, of \, ND_{56} = \frac{[ND_{56}]}{[ND_{56}] + [ND_{28}]} \, ,$$

where the square brackets denote the estimated concentration. The mean and standard deviation of this relative percentage was calculated from the 15 measurements for each of the 11 concentrations. Linear regression was used to determine the linearity of the estimate versus ground truth and to construct a calibration curve.

Finally, a second set of phantoms containing 0%, 25%, 50%, 75%, or 100% $ND_{56}$ were constructed and imaged using the same sequence. A total of 5 sets of images per phantom were acquired. The relative $ND_{56}$ concentration was calculated as before, but then scaled by the linear fit from the previous set of experiments to arrive at a final value. All processing was done in a pixel-by-pixel manner so that an image of the relative $ND_{56}$ concentration could be constructed.

AUTHOR INFORMATION

**Corresponding Author**

*Address correspondence to geoffrey.p.luke@dartmouth.edu

**Author Contributions**

The manuscript was written through contributions of all authors. All authors have given approval to the final version of the manuscript.

**Funding Sources**

This work was funded by the National Institutes of Health (R56DE033175) and a Prouty Pilot Grant from the Dartmouth Cancer Center.

ABBREVIATIONS

DLS, dynamic light scattering; DPPC, 1,2-dipalmitoyl-sn- glycero-3-phosphocholine; DSPE-PEG, 1,2- distearoyl-sn- glycero-3- phosphoethanolamine -N- (amino(polyethylene glycol)-2000) ; FUS, focused ultrasound; $ND_{28}$, perfluoropentane nanodroplets; $ND_{56}$ Perfluorohexane nanodroplets; PFC, perfluorocarbon; US, ultrasound;